\documentclass[reprint,showpacs,twocolumn,superscriptaddress,aps,prx]{revtex4-1}
\usepackage[utf8]{inputenc}
\usepackage[american,british]{babel}
\usepackage[T1]{fontenc}
\usepackage[pdftex]{graphicx}  
\usepackage{graphicx, xcolor}
\usepackage{dcolumn}
\usepackage{braket}
\usepackage{bm}
\usepackage{amsmath,amsthm,amssymb}
\usepackage{color}
\usepackage{verbatim}
\usepackage{ulem}
\usepackage[export]{adjustbox}

\definecolor{darkGreen}{RGB}{0,110,0}
\definecolor{darkBlue}{RGB}{0,0,130}
\usepackage{hyperref}

\hypersetup{
 colorlinks=true,
 linkcolor=blue,
 anchorcolor = blue,
 citecolor = blue,
 filecolor = blue,
 urlcolor = blue
}

\newcommand{\ee}{\mathrm{e}}
\usepackage{dsfont}
\usepackage{physics}
\def \be {\begin{equation}} 
\def \ee {\end{equation}} 
\def \l {\left(} 
\def \r {\right)} 
\def \la {\langle} 
\def \ra {\rangle}

\begin{document}
\title{A universal formula for the entanglement asymmetry of matrix product states}
\date{\today}

 \author{Luca Capizzi}
 \affiliation{SISSA, via Bonomea 265, 34136 Trieste, Italy}
 \affiliation{INFN Sezione di Trieste, via Bonomea 265, 34136 Trieste, Italy}
 \affiliation{Universit\'e Paris-Saclay, CNRS, LPTMS, 91405, Orsay, France.}

 \author{Vittorio Vitale}
 \affiliation{Univ. Grenoble Alpes, CNRS, LPMMC, 38000 Grenoble, France}

\begin{abstract}
Symmetry breaking is a fundamental concept in understanding quantum phases of matter, studied so far mostly through the lens of local order parameters.
Recently, a new entanglement-based probe of symmetry breaking has been introduced under the name of \textit{entanglement asymmetry}, which has been employed to investigate the mechanism of dynamical symmetry restoration. Here, we provide a universal formula for the entanglement asymmetry of matrix product states with finite bond dimension, valid in the large volume limit. We show that the entanglement asymmetry of any compact -- discrete or continuous -- group depends only on the symmetry breaking pattern, and is not related to any other microscopic features.
\end{abstract} 

\maketitle

In our current understanding of many-body quantum systems, symmetry and entanglement stand as two pivotal concepts, playing a crucial role in shaping phases of matter and characterizing quantum dynamics \cite{Montvay1994,Zeng19,sachdev_2000,Amico2002,ccd-09,l-15}. Surprisingly, their connection has received limited attention until fairly recently, when the concept of `symmetry-resolved entanglement' has been introduced~\cite{Casini2014Remarks,Xavier2018Equipartition,buividovich2008entanglement,laflorencie2014spin,Goldstein2018}. Since then, this interplay has been extensively investigated both theoretically~\cite{MurcianoDiGiulio2020,fraenkel2020symmetry,Tan2020Particle,pbc-21,capizzi2020symmetry} and experimentally~\cite{vitale2021symmetryresolved,rath2023entanglement}, and it has turned out to be crucial to fully understand some features of entanglement dynamics~\cite{Xavier2018Equipartition,vitale2021symmetryresolved} and detection~\cite{neven2021symmetryresolved}. However, the connection between \textit{symmetry breaking} and entanglement has remained elusive. Recently, such a connection has been explored by means of the \textit{entanglement asymmetry}. The latter has been used to analyse the restoration (or lack thereof) of a $U(1)$ symmetry in the quench dynamics of quantum spin chains~\cite{amc-23,amvc-23,bkccr-23}, and employed to characterize the symmetry-breaking pattern in field theory~\cite{CapizziMazzoni}.  In this work, we prove a general conjecture about the entanglement asymmetry, proposed in Ref.~\cite{CapizziMazzoni} for finite groups, and further extend it here to compact Lie groups. In particular, we show that the entanglement asymmetry of a large region is only related to the symmetry of the state, and does not rely on any additional features of the latter. We provide an extensive characterization for translational invariant Matrix Product States (MPS) \cite{fnw-92,Vidal-03} in the thermodynamic limit, supporting our theoretical results with numerical simulations using iDMRG~\cite{mcculloch2008infinite}.

\section{Introduction} 

We first summarize the main definitions, following closely Ref. \cite{CapizziMazzoni}, valid for the entanglement asymmetry of any compact group.

Let us consider a (possibly mixed) state $\rho$ of a bipartite system $A\cup \bar{A}$, described by the Hilbert space $
\mathcal{H} = \mathcal{H}_A \otimes \mathcal{H}_{\bar{A}}\,.
$
We assume that a finite group $G$ acts unitarily on $\mathcal{H}$ as map $G \ni g \mapsto \hat{g} \in \text{End}(\mathcal{H})$, with $\hat{g} = \hat{g}_A \otimes \hat{g}_{\bar{A}}$. Given $\rho$, we trace out the degrees of freedom of $\bar{A}$ and we construct the reduced density of matrix of $A$ as $\rho_A \equiv \text{Tr}_{\bar{A}}\l \rho \r$.
What we ask is whether $\rho_A$ is symmetric under the group $G$, namely whether $
\rho_A = \hat{g}_A \rho_A \hat{g}^{-1}_A$  holds for any $g \in G$, or is violated for some group elements, thus signaling a breaking of the symmetry. 
To do so, for finite groups we introduce the symmetrized state
\be\label{eq:rho_tilde}
\tilde{\rho}_A \equiv \frac{1}{|G|}\sum_{g \in G} \hat{g}_A\rho_A\hat{g}^{-1}_A\,
\ee
that can be generalized to
\be\label{eq:renyiasymmetry}
\tilde{\rho}_A \equiv \int_G \mathrm{d}g \,  \hat{g}_A\rho_A\hat{g}^{-1}_A\,
\ee
for generic compact Lie groups, with $\int_G dg$ the normalized Haar measure \cite{Vinberg-89}. It is easy to check that $\tilde{\rho}_A$ is symmetric under $G$ for any group element. Therefore, comparing the two states $\rho_A$ and $\tilde{\rho}_A$ would lead naturally to probing (spontaneous or explicit) symmetry-breaking at the level of the subsystem $A$. In doing so, one introduces the \textit{R\'enyi entanglement asymmetry}, defined as
\be\label{eq:Ent_asymm}
\Delta S_n \equiv \frac{1}{1-n}\log\frac{\text{Tr}\l \tilde{\rho}^n_A\r}{\text{Tr}\l \rho^n_A\r},
\ee
that is the R\'enyi entropy difference of the two states. As explained in Ref.~\cite{CapizziMazzoni} the computation of the R\'enyi entanglement asymmetry for $n\geq 2$ integer requires the calculation of the charged moments of $\rho_A$, that are the elements appearing in the sum
\be
\label{eq:tilderhoA}
\begin{aligned}
&\text{Tr}({\tilde{\rho}}^n_A ) =\\
&=\frac{1}{|G|^{n-1}}\sum_{g_i\in G} \text{Tr}(\rho_A g_1 \cdots \rho_A g_{n-1}\rho_A (g_1\cdots g_{n-1})^{-1} ).
\end{aligned}
\ee
Here, and later (if not specified explicitly), we omit the index $A$ and the hat on the group operators for notational convenience.

In this work, we are interested in the case where the quantum state $\rho$ is symmetric under a subgroup $H$ of $G$, defined as
\be\label{eq:Sym_H}
H \equiv  \{ h \in G | \rho = h\rho h^{-1}\} \subset G.
\ee
In this case, we say that the symmetry-breaking pattern $G\rightarrow H$, arises for the state $\rho$. The first main result of this work is to prove that the R\'enyi asymmetry of a very large subsystem $A$ is
\be\label{eq:conj}
\Delta S_n \simeq \log \frac{|G|}{|H|},
\ee
and thus it only depends on the symmetry-breaking pattern. Eq.~\eqref{eq:conj} appeared firstly as a conjecture in Ref. \cite{CapizziMazzoni}, and it has been previously proven only for the trivial case of zero-entanglement states (see also Ref. \cite{fac-23} for some specific cases).

The second contribution to the topic is the investigation of the entanglement asymmetry for generic compact Lie groups, whereas only the case of $U(1)$ has been previously considered explicitly~\cite{amc-23,amvc-23}. We prove that
\be\label{eq:Lie_asym}
\Delta S_n \simeq \frac{1}{2}\l\text{dim}\mathfrak{g}-\text{dim}\mathfrak{h}\r\log |A| + \dots,
\ee
with $\mathfrak{g},\mathfrak{h}$ the Lie algebras of $G,H$ respectively. This result is in agreement with the scaling $\Delta S_n \simeq \frac{1}{2} \log |A|$ observed in Refs. \cite{amc-23,fac-23} for the case of a broken $U(1)$ symmetry. Also, Eq.~\eqref{eq:Lie_asym} is reminiscent of the logarithmic scaling of entropy previously found in Refs. \cite{cd-11,cd-12,cd-13} for highly degenerate states.

Finally, we provide technical details regarding the explicit construction of the symmetrized state in terms of the symmetry sectors (see Appendix \ref{app:symm}). In this way, we make a connection with the original definition of the asymmetry proposed for abelian groups in Ref.~\cite{amc-23}, showing its natural extension to non-abelian groups.

\section{Entanglement asymmetry of finite groups }\label{sec:fin_groups}

\begin{figure}
    \includegraphics[width=\linewidth]{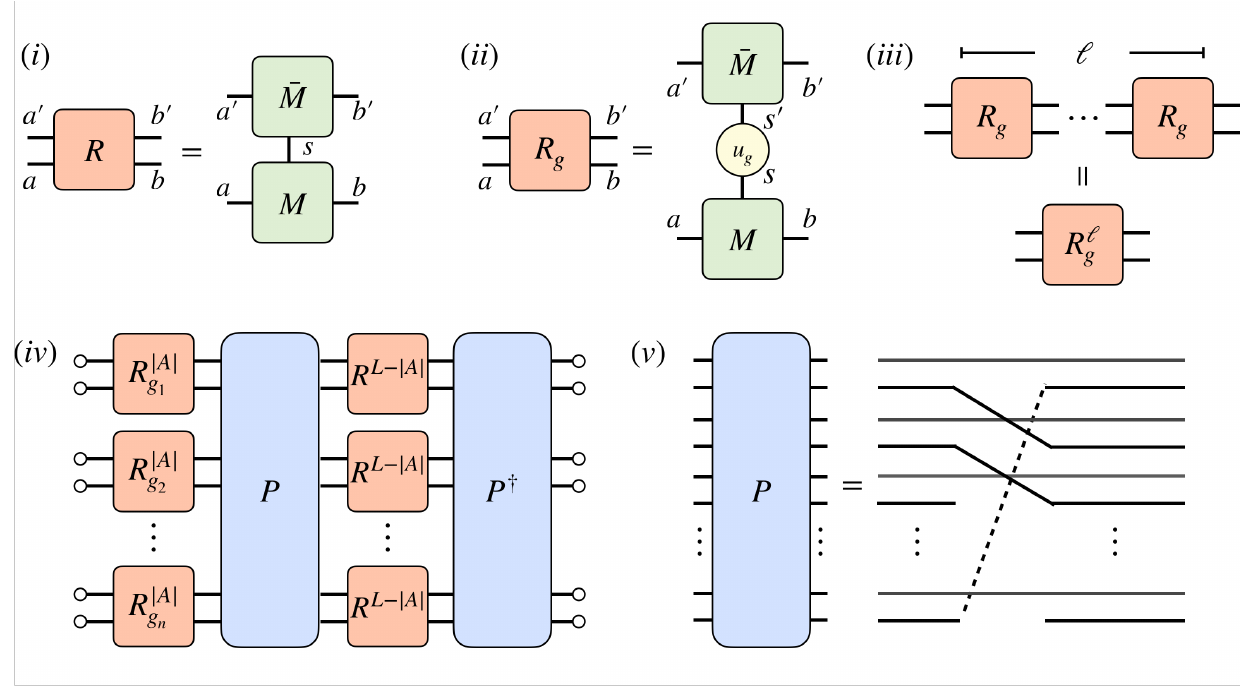}
        \caption{Diagrammatic representation of $(i)$ $R_{(a,a')(b,b')}$ as in Eq.~\eqref{eq:Rdef}, $(ii)$ $(R_g)_{(a,a')(b,b')}$ as in Eq.~\eqref{eq:R_g}, $(iii)$ $R^{\ell}_{(a,a')(b,b')}$, and $(iv)$ of the charged moments as in Eq.~\eqref{eq:tilderhoA} where $P$ in $(v)$ is the projector in Eq.~\eqref{eq:Pdef}. We replace the standard trace loop by circles at the end points that virtually connect to each other only horizontally at the same level.}
    \label{fig:sketch}
\end{figure}

We focus on translational invariant states with low entanglement, assuming that those can be efficiently described by MPS \cite{vc-06,Hastings-97}.

Let us consider a translational invariant (not normalized) MPS on a finite chain of size $L$~\cite{pwsvc-08}:
\be\label{eq:PsiL}
\ket{\Psi_L} = \sum^{d}_{s_1,\dots,s_L}\text{Tr}\l M_{s_1}\dots M_{s_L}\r \ket{s_1,\dots,s_L}.
\ee
Here $\{M_s\}$ are $D\times D$ matrices, and $D$ is the bond dimension of the MPS, and $d$ the dimension of the local Hilbert space. We are interested in the limit $L\rightarrow \infty$, such that the finite size effects are washed out; we denote the corresponding state by $\ket{\Psi}$.

The action of a global symmetry $G$ is implemented by the operator
\be
g = u_g \otimes \dots \otimes u_g,
\ee
with $u_g$ a $d\times d$ unitary matrix. By definition, $\ket{\Psi}$ is symmetric under the element $g$ whenever
\be
|\bra{\Psi}g\ket{\Psi}| \equiv \underset{L\rightarrow \infty}{\lim} \frac{|\bra{\Psi_L} g\ket{\Psi_L}|}{\bra{\Psi_L}\ket{\Psi_L}} =1,
\ee
otherwise, say if $|\bra{\Psi}g\ket{\Psi}| <1$, $\ket{\Psi}$ is said to be \textit{asymmetric} under $g$. We now discuss the consequence of the symmetry, or its lack, at the level of the local tensor $M$, relating them eventually to the large-scale behavior of the charged moments.

To do so, we first express
\be
\bra{\Psi_L}\ket{\Psi_L} = \text{Tr}\l R^L\r,
\ee
where $R$ is a $D^2\times D^2$ matrix given by
\be \label{eq:Rdef}
R_{(a,a')(b,b')} = \sum_s (M_s)_{a,b}\overline{(M_s)}_{a',b'},
\ee
with $a,b,a',b' = 1,\dots,D$. We represent it pictorially in Fig.~\ref{fig:sketch}$(i)$. We assume that $R$ has a single maximum eigenvalue in absolute value, a technical assumption that is physically equivalent to clustering of correlation functions as explained in \cite{SCHOLLWOCK201196} (see Appendix \ref{app:clust_hyp}). Furthermore, we normalize $M$ such that the maximum eigenvalue of $R$ is $1$. Within, this assumption, it is easy to show that $\ket{\Psi_L}$ is normalized in the infinite volume limit, namely
\be
\underset{L\rightarrow \infty}{\lim} 
 \bra{\Psi_L}\ket{\Psi_L} =1.
\ee

Similarly, we express
\be
 \bra{\Psi_L}g\ket{\Psi_L} = \text{Tr}\l R^L_g\r,
\ee
with $R_g$ a $D^2\times D^2$ matrix defined by
\be\label{eq:R_g}
\l R_g\r_{(a,a')(b,b')} = \sum_{s,s'} (M_s)_{a,b}\overline{(M_{s'})}_{a',b'}(u_g)_{s's}.
\ee
This is represented in Fig.~\ref{fig:sketch}$(ii)$. The bar denotes the complex conjugation.
As a consequence of $|\bra{\Psi_L}g\ket{\Psi_L}| \leq |\bra{\Psi_L}\ket{\Psi_L}|$, coming from the unitarity of $g$, it is easy to bound the spectrum of $R_g$ from above, for instance its operator norm satisfies
\be
\| R_g\| \equiv \underset{\ket{v}\neq 0} {\text{sup}}\frac{|\bra{v}R_g\ket{v}|}{\bra{v}\ket{v}} \leq 1.
\ee
In particular, $\| R_g\| = 1$ iff the state is symmetric since $|\bra{\Psi}g\ket{\Psi}| = \underset{L\rightarrow \infty}{\lim} |\text{Tr}\l R_g^L\r|$. Also, for the asymmetric case $\| R_g\| < 1$, one gets an exponential decay of the overlap
\be
\bra{\Psi_L} g\ket{\Psi_L} \sim \| R_g\|^L
\ee
in the large $L$ limit. 

At this point, we have all the ingredients to compute the charged moments of a subregion $A$. For the sake of simplicity, we focus on $A$ as a large but finite interval of length $|A|$. In this way, we express the moment of $\rho_A$ as 
\be\label{eq:Renyi_MPS}
\text{Tr}\l \rho_A^n\r = \underset{L\rightarrow \infty}{\lim} \text{Tr}\l \l R^{\otimes{n}}\r^{|A|} P\l R^{\otimes{n}}\r^{L-|A|}P^\dagger \r,
\ee
where $P$ is the $D^{2n}\times D^{2n}$ matrix associated with the following permutation of bonds (see e.g. Ref. \cite{bkccr-23})
\be\label{eq:Pdef}
\begin{pmatrix}
1 & 2 & 3 &\dots  & 2n-1 & 2n \\
1 & 4 & 3 & \dots & 2n-1 & 2 \\
\end{pmatrix}
\ee
as represented pictorially in Fig.~\ref{fig:sketch}$(v)$. Let us observe that, since $\|R\|=1$, one has
\be
\underset{L\rightarrow \infty}{\lim} R^L = \Pi,
\ee
with $\Pi$ a rank-one projector, namely $\Pi^2=\Pi$ and $\text{Tr}(\Pi)=1$. As an important consequence, in the limit $|A|\rightarrow \infty$ the R\'enyi entropy converges (and it satisfies the area law~\cite{SCHOLLWOCK201196}) to
\be\label{eq:area_law}
S_n = \frac{1}{1-n}\log \text{Tr}\l \Pi^{\otimes n} P \Pi^{\otimes n} P^\dagger \r.
\ee

With a similar approach, we can compute the charged moments of $A$ as
\be
\begin{split}
&\text{Tr}\l \rho_A g_{1}\dots \rho_A g_{n}\r =\\
& \underset{L\rightarrow \infty}{\lim} \text{Tr}\l \l R_{g_1}\otimes \cdots \otimes R_{g_n}\r^{|A|} P\l R^{\otimes{n}}\r^{L-|A|}P^\dagger \r
\end{split}.
\ee
From the expression above, it is clear that the charged moments can only converge to a constant or vanish exponentially in the large $|A|$ limit, depending on the operator norm of $\{R_{g_j}\}_{j=1,\dots,n}$. For instance, if $g_j$ is a symmetry of the state, say $g_j \in H$, then $\| R_{g_j}\| =1$. Thus, given $e^{i\phi_j}$ the largest eigenvalue (in absolute value) of $R_{g_j}$ we have
\be
\underset{L\rightarrow \infty}{\lim} R^{L}_{g_j} e^{-i\phi_j L} = \Pi_j,
\ee
with $\Pi_j$ a rank-one projector, leading immediately to
\be
\label{eq:area_law1}
\begin{split}
&\underset{|A|\rightarrow \infty}{\lim} \text{Tr}\l \rho_A g_{1}\dots \rho_A g_{n}\r e^{-i(\phi_1+\dots+\phi_n)|A|}= \\
&\text{Tr}\l (\Pi_1\otimes\dots \otimes \Pi_n) P \Pi^{\otimes n} P^\dagger\r.
\end{split}
\ee
On the contrary, whenever $\|R_{g_j}\|<1$ for some $j$, then an exponential decay is observed and one gets
\be\label{eq:ch_mom_decay}
\text{Tr}\l \rho_A g_{1}\dots \rho_A g_{n}\r \sim \l \|R_{g_1}\| \dots \|R_{g_1}\| \r^{|A|}.
\ee
We finally notice that the terms in Eq.~\eqref{eq:tilderhoA} with $g_i \in H$ are equal to $\text{Tr}(\rho^n_A)$, since, whenever $g \in H$ one has $[\rho_A,g]=0$, as a consequence of the definition Eq.~\eqref{eq:Sym_H} (see Ref. \cite{CapizziMazzoni}). Thanks to the last property and the vanishing of the other charged moments in Eq.\eqref{eq:ch_mom_decay}, one gets from Eq.~\eqref{eq:tilderhoA} the universal prediction in Eq.~\eqref{eq:conj} for finite groups.

To conclude this section, we emphasize that we did not find any direct relation between the correlation length of the state $\ket{\Psi}$, related to the spectral gap of $R$, and the rate of exponential decay of the charged moments, ruled instead by the norm of $R_g$ (see Appendix \ref{app:Area_law}). This might seem in contrast with the previous analysis of the massive Ising field theory in the ordered phase of Ref. \cite{CapizziMazzoni}, where a unique length $\xi = (2m)^{-1}$ is present, and the two quantities mentioned above are closely related. However, we believe that the latter feature is specific to relativistic field theories, as it does not have a counterpart to lattice models.

\section{Generalization to compact Lie groups}\label{sec:Lie}

In this section, we prove the formula in  Eq.~\eqref{eq:Lie_asym} for compact Lie groups. In particular, while the analysis of the exponential decay of the charged moments in Sec.~\ref{sec:fin_groups} applies to both finite and continuous groups, differences arise in the computation of the entanglement asymmetry, depending on whether the Haar measure is discrete/continuous. We provide a saddle point analysis, generalizing the approach of Ref.~\cite{amc-23}, where the $U(1)$ group was considered, to tackle any compact Lie group $G$.

To do so, we first express the asymmetry as an integral
\be\label{eq:Integral_Haar}
\Delta S_n = \frac{1}{1-n}\log\l \int_{G} dg_1 \dots \int_{G} dg_{n-1} f(g_1,\dots,g_{n-1})\r,
\ee
with $f:G^{n-1}\rightarrow \mathbb{C}$ defined by
\be\label{eq:f_fun}
f(\mathbf{g}) \equiv \frac{\text{Tr}\l \rho_A g_{1}\rho_A g_2 \dots \rho_A (g_1\dots g_{n-1})^{-1}\r}{\text{Tr}\l \rho^n_A\r},
\ee
with $\mathbf{g} \equiv (g_1,\dots,g_{n-1})$.
From our analysis, we know that $|f(\mathbf{g})| \leq 1$, and $|f(\mathbf{g})| = 1$ iff $\mathbf{g} \in H^{n-1}$. Also, from the analysis of Sec. \ref{sec:fin_groups}, we have that $|f(\mathbf{g})|$ goes to zero exponentially in $|A|$ whenever $\mathbf{g} \notin H^{n-1}$. These considerations suggest that the dominant contribution to the integral in Eq.~\eqref{eq:Integral_Haar} comes from a small neighborhood of the submanifold $H^{n-1}$, and a saddle point analysis around those points can be performed. Some technical details regarding the validity of this approach are summarized in the Appendix \ref{app:saddle}.

To perform the integral in Eq.~\eqref{eq:Integral_Haar}, it is convenient to decompose $G$ in terms of $H$ and its coset $G/H$.
For the sake of simplicity, we assume explicitly that $G$ is connected and $G/H$ (that is not a group, in general) is a smooth manifold, albeit these hypotheses can be easily relaxed. Then, for any $g_j \in G$, we decompose
\be
g_j = h_j e^{X_j},
\ee
with $h_j \in H$ and $X_j \in \mathfrak{g}/\mathfrak{h}$, that is the tangent space of the coset $G/H$. In a neighborhood of $H^{n-1}$ we expand at second order the function $f$ (see Appendix \ref{app:saddle})
\be\label{eq:f_2ord}
\log f(\mathbf{g}) \simeq  -\frac{1}{2}\mathbf{X}^T N(\mathbf{h})\mathbf{X} |A|,
\ee
with $\mathbf{X}^T N(\mathbf{h})\mathbf{X}$ a non-degenerate quadratic form over $(\mathfrak{g}/\mathfrak{h})^{n-1}$ depending on $\mathbf{h} = (h_1,\dots,h_{n-1})$. Since $|f(\mathbf{g})|$ has a local maximum at $\mathbf{g} \in H^{n-1}$, corresponding to $\mathbf{X}=0$, the quadratic form above should be positive, and we can perform the Gaussian integral over $\mathbf{X}$ as follows
\be\label{eq:integ_f}
\begin{split}
&\int_{G^{n-1}}d\mathbf{g}  f(\mathbf{g}) \simeq \\
&\int_{H^{n-1}} d\mathbf{h} \int_{(\mathfrak{g}/\mathfrak{h})^{n-1}} d\mathbf{X} \exp\l  -\frac{1}{2}\mathbf{X}^T N(\mathbf{h})\mathbf{X} |A|\r \simeq \\
& \l \frac{|A|}{2\pi}\r^{- \frac{n-1}{2}\text{dim}(\mathfrak{g}/\mathfrak{h})}\int_{H^{n-1}} d\mathbf{h} \l\text{det} N (\mathbf{h})\r^{-1/2}.
\end{split}
\ee
In the end, we got a power-law decay in $|A|$, while the last integral over $\mathbf{h}$ gives just a model-dependent proportionality constant that is beyond our purpose. Putting everything together, we eventually get Eq.~\eqref{eq:Lie_asym} up to finite (order $O(1)$) terms in the limit $|A| \rightarrow \infty$.

\section{Numerical results}\label{sec:Numerics}
In this section, we discuss the ground state of a quantum spin chain, and we provide numerical results for its entanglement asymmetry to support our predictions. Let us consider the spin-1/2 XXZ model defined as
\be
H_{XXZ}=\sum_j \left(\sigma_j^x\sigma_{j+1}^{x}+\sigma_j^{y}\sigma_{j+1}^{y}\right)+\Delta \sum_j \sigma_j^z\sigma_{j+1}^z,
\ee
where $\sigma^{\alpha}_j$ ($\alpha=x,y,z$) are the Pauli matrices at position $j$. We focus on the discrete group $G$ generated by the rotation of $\pi/2$ around the $y$ axis, namely $\prod_je^{-i\frac{\theta}{2}\sigma^y_j}$ with $\theta=0,\pi/2,\pi,3\pi/2$. The Hamiltonian $H_{XXZ}$ is invariant under the rotations of $\theta=0,\pi$, while the angles $\theta=\pi/2,3\pi/2$ correspond to explicitly broken elements. We consider the ground state of the model for various parameters $\Delta$ and we aim to characterize the entanglement asymmetry of large regions: careful is needed since, by the Lieb-Schultz-Mattis Theorem\cite{lsm-61}, the ground state of the model is never gapped. We discuss separately the phases as a function of $\Delta$.

For $\Delta>1$, the antiferromagnetic phase, translational symmetry and the rotation of $\theta=\pi$ around $y$ are spontaneously broken, and two degenerate ground states are present in the infinite volume limit. These are adiabatically connected to the N\'eel states $\ket{\Psi_0} = \ket{\uparrow\downarrow\uparrow\downarrow,\dots},\ket{\downarrow\uparrow\downarrow\uparrow,\dots}$, which become exact ground states in the limit $\Delta\rightarrow +\infty$. At finite size, the two states above hybridize, and a single non-clustering ground state is present. However, we focus on the clustering state with finite correlation length obtained via tensor network methods (both DMRG and iDMRG in the following), starting from a N\'eel state; this is an approximation of the ground state in the infinite volume limit, for which the theory developed in the main text is expected to be predictive. In this case, the invariant subgroup $H \subset G$ associated to this state is the trivial group (rotations with $\theta=0$), and we get $\Delta S_2 \simeq \log \frac{|G|}{|H|} = \log 4$ for large subsystem sizes.

For $\Delta<-1$, the ferromagnetic phase, the two states $\ket{\Psi_0} =\ket{\uparrow\uparrow\dots},\ket{\downarrow\downarrow\dots}$ are exact ground states at finite size. We study the ground states with all spin up along the $z$ axis, which does not have spatial entanglement. For the state above, an elementary calculation gives the charged moments of an interval of length $\ell$ as
\be
\text{Tr}\l \rho_A g \rho_A g^{-1}\r =
\begin{cases}
1 \quad \theta=0,\\
\frac{1}{2^\ell} \quad  \theta = \pi/2, 3\pi/2,\\
0 \quad \theta = \pi.
\end{cases}
\ee
Therefore, the asymmetry of the group $G$ converges exponentially fast to $\Delta S_2 \simeq \log 4$ in the subsystem size. This is compatible with the general prediction of Eq.~\eqref{eq:conj}, since the ground state above is broken by the rotation with $\theta=\pi$ and it admits the trivial group as invariant subgroup.

For $\Delta = (-1,1)$ the system is critical, a single ground state is present at finite size and the gap decays algebraically in the infinite size limit. This regime is not well-described by the formalism presented in the main text, since critical states are known to be not efficiently represented by MPS due to their algebraically decaying correlations. Still, we conjecture the validity of the main prediction (Eq.~\eqref{eq:conj}), which reads as $\Delta S_2 \simeq \log 2$ since spontaneous symmetry breaking does not occur and the ground state shares the same symmetries of the Hamiltonian (in particular it is invariant under $\theta = \pi$). We finally observe that the limit $\Delta \rightarrow 1^{-}$ is tricky, since the hamiltonian becomes rotational invariant for any $\theta$; however, a comprehensive discussion of the pathological case $\Delta=1$ is beyond our scope.

\begin{figure}
\centering
    \includegraphics[width=\linewidth]{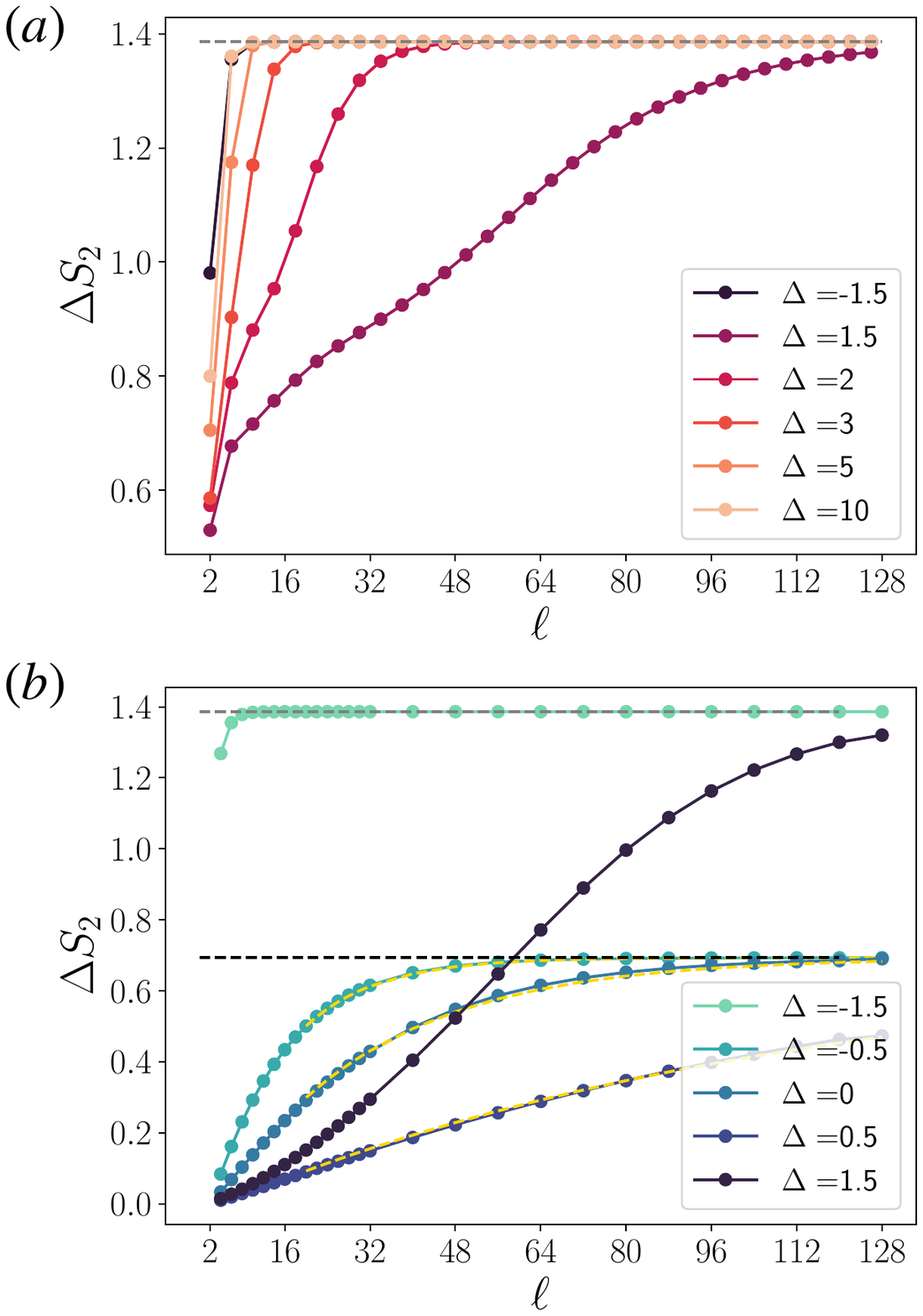}
    \caption{Rényi-2 entanglement asymmetry $\Delta S_2$ of the ground state of the XXZ model for several values of $\Delta$, as a function of the length of the system considered $\ell$, computed using $(a)$ iDMRG, bond dimension considered $D=128$ and $(b)$ DMRG, $D=160$. The dashed grey line is the asymptotic value $\log 4$, while the dashed black line marks the value $\log 2$. We highlight the exponential fit of $\Delta S_2$ in the critical region, by dashed yellow lines.}
        \label{fig:asymmetryXXZ}
\end{figure}

In Fig.~\ref{fig:asymmetryXXZ} we provide numerical results for all the phases employing both DMRG and iDMRG. The latter (Fig.~\ref{fig:asymmetryXXZ}$(a)$) is efficient in the antiferromagnetic and ferromagnetic phases since the correlation length is finite. 
We consider a bond dimension $D=128$ (which implies a truncation error $\sim 10^{-9}$) and a unit cell of 4 sites.
We observe that $\Delta S_2$ converges to $\log4$ as predicted before. The convergence is slower near the critical point
$\Delta  = 1$, and we believe that this is due both to the divergence of the correlation length and the restoration of the symmetry at $\Delta = 1$.

We further probe the critical regime $\Delta \in (-1,1)$ by means of DMRG. For instance, we pick a finite-size chain with open boundary conditions and we consider as a subsystem the entire chain. 
We set the bond dimension equal to $D=160$, which implies a truncation error $\sim 10^{-7}$, and we display the results in Fig. ~\ref{fig:asymmetryXXZ}$(b)$.
We observe that $\Delta S_2$ approaches exponentially $\log2$ as expected. We show the exponential fits with dashed yellow lines.
Furthermore, we check that in the gapped regime the results obtained at finite size via DMRG show similar features with those at infinite size (with iDRMG).

\section{Conclusions}

In this work, we establish the validity of a universal prediction of the entanglement asymmetry for both discrete (Eq.~\eqref{eq:conj}) and continuous groups (Eq.~\eqref{eq:Lie_asym}). Our analysis is based on the properties of translational invariant MPS with finite bond dimension, and our proof relies on the relation between the symmetry of the state and the spectral properties of the corresponding local tensors. Our results are compatible with the universal values of the entanglement asymmetry observed in Ref. \cite{fac-23} in a global quench at short times (compared to the subsystem size) and in Ref. \cite{CapizziMazzoni} for the ground state of the Ising field theory.

Our predictions mainly rely on the proof of the exponential decay in Eq. \eqref{eq:ch_mom_decay}, derived for one dimensional tensor network with finite bond dimension, which comes from the bound $||R_g||<1, g\notin H$ discussed in the main text.

Interesting generalizations could be provided. For example, we expect that, if boundary effects are present and symmetries are broken by the boundary conditions only (see e.g. \cite{bcp-23}), the corresponding charged moments should not decay exponentially to zero and violation to our universal predictions could be observed. Also, we believe that our analysis could be extended to mixed states described by Matrix Product Operator (MPO), which are known to efficiently simulate thermal states \cite{msvc-15} and possibly other stationary ensembles~\cite{dubail2017entanglement}.  Finally, tensor networks in higher dimensions, as the Projected Entangled-pair States (PEPS) \cite{vc-04}, important for topological phases of matter and lattice gauge theories, might be investigated as well within a similar framework.

Some important questions remain open. First, it is not clear what happens for ground states of critical hamiltonians, which are known to show algebraic decay of correlation functions, and therefore are not described by MPS \cite{SCHOLLWOCK201196} with finite bond-dimension.  In principle, this might not be an obstacle to the main result, as long as an MPS with infinite bond dimension describes the state and $||R_g||$ is strictly bounded by $1$ for $g$ not belonging to the invariant group $H$. In particular, a work on the full-counting statistics of the critical Ising chain \cite{Stephan-14} suggests that the generating function of the non-conserved charges and, more in general, the charged moments might still decay exponentially, even if the correlation length of the state diverges.

Also, some fundamental aspects of the dynamical restoration of symmetries seem to be missing. For instance, it would be interesting to understand a general criterion, besides the specific example in Ref.~\cite{amvc-23}, to detect the lack of symmetry restoration, relating the (possibly non-abelian) symmetries of the stationary states and the conserved charges.

Finally, let us observe that the entanglement asymmetry is also experimentally measurable by means of randomized measurements~\cite{elben2023randomized}, equivalently to symmetry-resolved entanglement~\cite{vitale2021symmetryresolved,neven2021symmetryresolved}.

\section*{Authors contribution}
L.C. devised the conjecture and proved it analytically. V.V. performed the numerical simulations. Both authors discussed the results and wrote the manuscript.

\section*{Acknowledgements}

Work in Grenoble is funded by the French National Research Agency via QUBITAF (ANR-22-PETQ-0004, Plan France 2030).
L.C. acknowledges support from ERC under Consolidator grant number 771536 (NEMO). L.C. is grateful to Olalla Castro-Alvaredo, Michele Fossati and Colin Rylands for valuable insights regarding the manuscript.
V.V. acknowledges useful discussions with Sara Murciano and Aniket Rath on related topics.

\bibliography{biblio}

\begin{thebibliography}{45}%
\makeatletter
\providecommand \@ifxundefined [1]{%
 \@ifx{#1\undefined}
}%
\providecommand \@ifnum [1]{%
 \ifnum #1\expandafter \@firstoftwo
 \else \expandafter \@secondoftwo
 \fi
}%
\providecommand \@ifx [1]{%
 \ifx #1\expandafter \@firstoftwo
 \else \expandafter \@secondoftwo
 \fi
}%
\providecommand \natexlab [1]{#1}%
\providecommand \enquote  [1]{``#1''}%
\providecommand \bibnamefont  [1]{#1}%
\providecommand \bibfnamefont [1]{#1}%
\providecommand \citenamefont [1]{#1}%
\providecommand \href@noop [0]{\@secondoftwo}%
\providecommand \href [0]{\begingroup \@sanitize@url \@href}%
\providecommand \@href[1]{\@@startlink{#1}\@@href}%
\providecommand \@@href[1]{\endgroup#1\@@endlink}%
\providecommand \@sanitize@url [0]{\catcode `\\12\catcode `\$12\catcode
  `\&12\catcode `\#12\catcode `\^12\catcode `\_12\catcode `\%12\relax}%
\providecommand \@@startlink[1]{}%
\providecommand \@@endlink[0]{}%
\providecommand \url  [0]{\begingroup\@sanitize@url \@url }%
\providecommand \@url [1]{\endgroup\@href {#1}{\urlprefix }}%
\providecommand \urlprefix  [0]{URL }%
\providecommand \Eprint [0]{\href }%
\providecommand \doibase [0]{http://dx.doi.org/}%
\providecommand \selectlanguage [0]{\@gobble}%
\providecommand \bibinfo  [0]{\@secondoftwo}%
\providecommand \bibfield  [0]{\@secondoftwo}%
\providecommand \translation [1]{[#1]}%
\providecommand \BibitemOpen [0]{}%
\providecommand \bibitemStop [0]{}%
\providecommand \bibitemNoStop [0]{.\EOS\space}%
\providecommand \EOS [0]{\spacefactor3000\relax}%
\providecommand \BibitemShut  [1]{\csname bibitem#1\endcsname}%
\let\auto@bib@innerbib\@empty
\bibitem [{\citenamefont {Montvay}\ and\ \citenamefont
  {Muenster}(1994)}]{Montvay1994}%
  \BibitemOpen
  \bibfield  {author} {\bibinfo {author} {\bibfnamefont {I.}~\bibnamefont
  {Montvay}}\ and\ \bibinfo {author} {\bibfnamefont {G.}~\bibnamefont
  {Muenster}},\ }\href {https://doi.org/10.1017/CBO9780511470783} {\emph
  {\bibinfo {title} {{Quantum Fields on a lattice}}}}\ (\bibinfo  {publisher}
  {Cambridge Univ. Press, Cambridge},\ \bibinfo {year} {1994})\BibitemShut
  {NoStop}%
\bibitem [{\citenamefont {Zeng}\ \emph {et~al.}(2019)\citenamefont {Zeng},
  \citenamefont {Chen}, \citenamefont {Zhou},\ and\ \citenamefont
  {Wen}}]{Zeng19}%
  \BibitemOpen
  \bibfield  {author} {\bibinfo {author} {\bibfnamefont {B.}~\bibnamefont
  {Zeng}}, \bibinfo {author} {\bibfnamefont {X.}~\bibnamefont {Chen}}, \bibinfo
  {author} {\bibfnamefont {D.-L.}\ \bibnamefont {Zhou}}, \ and\ \bibinfo
  {author} {\bibfnamefont {X.-G.}\ \bibnamefont {Wen}},\ }\href
  {https://doi.org/10.1007/978-1-4939-9084-9} {\emph {\bibinfo {title} {Quantum
  Information Meets Quantum Matter}}}\ (\bibinfo  {publisher} {Springer New
  York},\ \bibinfo {year} {2019})\BibitemShut {NoStop}%
\bibitem [{\citenamefont {Sachdev}(2000)}]{sachdev_2000}%
  \BibitemOpen
  \bibfield  {author} {\bibinfo {author} {\bibfnamefont {S.}~\bibnamefont
  {Sachdev}},\ }\href {https://doi.org/10.1017/CBO9780511622540} {\emph
  {\bibinfo {title} {Quantum Phase Transitions}}}\ (\bibinfo  {publisher}
  {Cambridge University Press},\ \bibinfo {year} {2000})\BibitemShut {NoStop}%
\bibitem [{\citenamefont {Amico}\ \emph {et~al.}(2008)\citenamefont {Amico},
  \citenamefont {Fazio}, \citenamefont {Osterloh},\ and\ \citenamefont
  {Vedral}}]{Amico2002}%
  \BibitemOpen
  \bibfield  {author} {\bibinfo {author} {\bibfnamefont {L.}~\bibnamefont
  {Amico}}, \bibinfo {author} {\bibfnamefont {R.}~\bibnamefont {Fazio}},
  \bibinfo {author} {\bibfnamefont {A.}~\bibnamefont {Osterloh}}, \ and\
  \bibinfo {author} {\bibfnamefont {V.}~\bibnamefont {Vedral}},\ }\href
  {\doibase 10.1103/RevModPhys.80.517} {\bibfield  {journal} {\bibinfo
  {journal} {Review of Modern Physics}\ }\textbf {\bibinfo {volume} {80}},\
  \bibinfo {pages} {517} (\bibinfo {year} {2008})}\BibitemShut {NoStop}%
\bibitem [{\citenamefont {Calabrese}\ \emph {et~al.}(2009)\citenamefont
  {Calabrese}, \citenamefont {Cardy},\ and\ \citenamefont {Doyon}}]{ccd-09}%
  \BibitemOpen
  \bibfield  {author} {\bibinfo {author} {\bibfnamefont {P.}~\bibnamefont
  {Calabrese}}, \bibinfo {author} {\bibfnamefont {J.}~\bibnamefont {Cardy}}, \
  and\ \bibinfo {author} {\bibfnamefont {B.}~\bibnamefont {Doyon}},\ }\href
  {https://doi.org/10.1088/1751-8121/42/50/500301} {\bibfield  {journal}
  {\bibinfo  {journal} {Journal of Physics A: Mathematical and Theoretical}\
  }\textbf {\bibinfo {volume} {42}},\ \bibinfo {pages} {500301} (\bibinfo
  {year} {2009})}\BibitemShut {NoStop}%
\bibitem [{\citenamefont {Laflorencie}(2016)}]{l-15}%
  \BibitemOpen
  \bibfield  {author} {\bibinfo {author} {\bibfnamefont {N.}~\bibnamefont
  {Laflorencie}},\ }\href {\doibase 10.1016/j.physrep.2016.06.008} {\bibfield
  {journal} {\bibinfo  {journal} {Phys. Rep.}\ }\textbf {\bibinfo {volume}
  {646}},\ \bibinfo {pages} {1 } (\bibinfo {year} {2016})}\BibitemShut
  {NoStop}%
\bibitem [{\citenamefont {Casini}\ \emph {et~al.}(2014)\citenamefont {Casini},
  \citenamefont {Huerta},\ and\ \citenamefont {Rosabal}}]{Casini2014Remarks}%
  \BibitemOpen
  \bibfield  {author} {\bibinfo {author} {\bibfnamefont {H.}~\bibnamefont
  {Casini}}, \bibinfo {author} {\bibfnamefont {M.}~\bibnamefont {Huerta}}, \
  and\ \bibinfo {author} {\bibfnamefont {J.~A.}\ \bibnamefont {Rosabal}},\
  }\href {\doibase 10.1103/PhysRevD.89.085012} {\bibfield  {journal} {\bibinfo
  {journal} {Physical Review D}\ }\textbf {\bibinfo {volume} {89}},\ \bibinfo
  {pages} {085012} (\bibinfo {year} {2014})}\BibitemShut {NoStop}%
\bibitem [{\citenamefont {Xavier}\ \emph {et~al.}(2018)\citenamefont {Xavier},
  \citenamefont {Alcaraz},\ and\ \citenamefont
  {Sierra}}]{Xavier2018Equipartition}%
  \BibitemOpen
  \bibfield  {author} {\bibinfo {author} {\bibfnamefont {J.~C.}\ \bibnamefont
  {Xavier}}, \bibinfo {author} {\bibfnamefont {F.~C.}\ \bibnamefont {Alcaraz}},
  \ and\ \bibinfo {author} {\bibfnamefont {G.}~\bibnamefont {Sierra}},\ }\href
  {\doibase 10.1103/PhysRevB.98.041106} {\bibfield  {journal} {\bibinfo
  {journal} {Physical Review B}\ }\textbf {\bibinfo {volume} {98}},\ \bibinfo
  {pages} {041106} (\bibinfo {year} {2018})}\BibitemShut {NoStop}%
\bibitem [{\citenamefont {Buividovich}\ and\ \citenamefont
  {Polikarpov}(2008)}]{buividovich2008entanglement}%
  \BibitemOpen
  \bibfield  {author} {\bibinfo {author} {\bibfnamefont {P.}~\bibnamefont
  {Buividovich}}\ and\ \bibinfo {author} {\bibfnamefont {M.}~\bibnamefont
  {Polikarpov}},\ }\href
  {https://www.sciencedirect.com/science/article/abs/pii/S0370269308012987?via%3Dihub}
  {\bibfield  {journal} {\bibinfo  {journal} {Physics Letters B}\ }\textbf
  {\bibinfo {volume} {670}},\ \bibinfo {pages} {141} (\bibinfo {year}
  {2008})}\BibitemShut {NoStop}%
\bibitem [{\citenamefont {Laflorencie}\ and\ \citenamefont
  {Rachel}(2014)}]{laflorencie2014spin}%
  \BibitemOpen
  \bibfield  {author} {\bibinfo {author} {\bibfnamefont {N.}~\bibnamefont
  {Laflorencie}}\ and\ \bibinfo {author} {\bibfnamefont {S.}~\bibnamefont
  {Rachel}},\ }\href
  {https://iopscience.iop.org/article/10.1088/1742-5468/2014/11/P11013}
  {\bibfield  {journal} {\bibinfo  {journal} {Journal of Statistical Mechanics:
  Theory and Experiment}\ }\textbf {\bibinfo {volume} {2014}},\ \bibinfo
  {pages} {P11013} (\bibinfo {year} {2014})}\BibitemShut {NoStop}%
\bibitem [{\citenamefont {Goldstein}\ and\ \citenamefont
  {Sela}(2018)}]{Goldstein2018}%
  \BibitemOpen
  \bibfield  {author} {\bibinfo {author} {\bibfnamefont {M.}~\bibnamefont
  {Goldstein}}\ and\ \bibinfo {author} {\bibfnamefont {E.}~\bibnamefont
  {Sela}},\ }\href {\doibase 10.1103/PhysRevLetters120.200602} {\bibfield
  {journal} {\bibinfo  {journal} {Physical Review Letters}\ }\textbf {\bibinfo
  {volume} {120}},\ \bibinfo {pages} {200602} (\bibinfo {year}
  {2018})}\BibitemShut {NoStop}%
\bibitem [{\citenamefont {Murciano}\ \emph {et~al.}(2020)\citenamefont
  {Murciano}, \citenamefont {Giulio},\ and\ \citenamefont
  {Calabrese}}]{MurcianoDiGiulio2020}%
  \BibitemOpen
  \bibfield  {author} {\bibinfo {author} {\bibfnamefont {S.}~\bibnamefont
  {Murciano}}, \bibinfo {author} {\bibfnamefont {G.~D.}\ \bibnamefont
  {Giulio}}, \ and\ \bibinfo {author} {\bibfnamefont {P.}~\bibnamefont
  {Calabrese}},\ }\href {https://doi.org/10.1007%2Fjhep08%282020%29073}
  {\bibfield  {journal} {\bibinfo  {journal} {Journal of High Energy Physics}\
  }\textbf {\bibinfo {volume} {2020}} (\bibinfo {year} {2020})}\BibitemShut
  {NoStop}%
\bibitem [{\citenamefont {Fraenkel}\ and\ \citenamefont
  {Goldstein}(2020)}]{fraenkel2020symmetry}%
  \BibitemOpen
  \bibfield  {author} {\bibinfo {author} {\bibfnamefont {S.}~\bibnamefont
  {Fraenkel}}\ and\ \bibinfo {author} {\bibfnamefont {M.}~\bibnamefont
  {Goldstein}},\ }\href
  {https://iopscience.iop.org/article/10.1088/1742-5468/ab7753} {\bibfield
  {journal} {\bibinfo  {journal} {Journal of Statistical Mechanics: Theory and
  Experiment}\ }\textbf {\bibinfo {volume} {2020}},\ \bibinfo {pages} {033106}
  (\bibinfo {year} {2020})}\BibitemShut {NoStop}%
\bibitem [{\citenamefont {Tan}\ and\ \citenamefont
  {Ryu}(2020)}]{Tan2020Particle}%
  \BibitemOpen
  \bibfield  {author} {\bibinfo {author} {\bibfnamefont {M.~T.}\ \bibnamefont
  {Tan}}\ and\ \bibinfo {author} {\bibfnamefont {S.}~\bibnamefont {Ryu}},\
  }\href {\doibase 10.1103/PhysRevB.101.235169} {\bibfield  {journal} {\bibinfo
   {journal} {Physical Review B}\ }\textbf {\bibinfo {volume} {101}},\ \bibinfo
  {pages} {235169} (\bibinfo {year} {2020})}\BibitemShut {NoStop}%
\bibitem [{\citenamefont {Parez}\ \emph {et~al.}(2021)\citenamefont {Parez},
  \citenamefont {Bonsignori},\ and\ \citenamefont {Calabrese}}]{pbc-21}%
  \BibitemOpen
  \bibfield  {author} {\bibinfo {author} {\bibfnamefont {G.}~\bibnamefont
  {Parez}}, \bibinfo {author} {\bibfnamefont {R.}~\bibnamefont {Bonsignori}}, \
  and\ \bibinfo {author} {\bibfnamefont {P.}~\bibnamefont {Calabrese}},\ }\href
  {https://journals.aps.org/prb/abstract/10.1103/PhysRevB.103.L041104}
  {\bibfield  {journal} {\bibinfo  {journal} {Physical Review B}\ }\textbf
  {\bibinfo {volume} {103}},\ \bibinfo {pages} {L041104} (\bibinfo {year}
  {2021})}\BibitemShut {NoStop}%
\bibitem [{\citenamefont {Capizzi}\ \emph {et~al.}(2020)\citenamefont
  {Capizzi}, \citenamefont {Ruggiero},\ and\ \citenamefont
  {Calabrese}}]{capizzi2020symmetry}%
  \BibitemOpen
  \bibfield  {author} {\bibinfo {author} {\bibfnamefont {L.}~\bibnamefont
  {Capizzi}}, \bibinfo {author} {\bibfnamefont {P.}~\bibnamefont {Ruggiero}}, \
  and\ \bibinfo {author} {\bibfnamefont {P.}~\bibnamefont {Calabrese}},\ }\href
  {https://iopscience.iop.org/article/10.1088/1742-5468/ab96b6/meta} {\bibfield
   {journal} {\bibinfo  {journal} {Journal of Statistical Mechanics: Theory and
  Experiment}\ }\textbf {\bibinfo {volume} {2020}},\ \bibinfo {pages} {073101}
  (\bibinfo {year} {2020})}\BibitemShut {NoStop}%
\bibitem [{\citenamefont {Vitale}\ \emph {et~al.}(2022)\citenamefont {Vitale},
  \citenamefont {Elben}, \citenamefont {Kueng}, \citenamefont {Neven},
  \citenamefont {Carrasco}, \citenamefont {Kraus}, \citenamefont {Zoller},
  \citenamefont {Calabrese}, \citenamefont {Vermersch},\ and\ \citenamefont
  {Dalmonte}}]{vitale2021symmetryresolved}%
  \BibitemOpen
  \bibfield  {author} {\bibinfo {author} {\bibfnamefont {V.}~\bibnamefont
  {Vitale}}, \bibinfo {author} {\bibfnamefont {A.}~\bibnamefont {Elben}},
  \bibinfo {author} {\bibfnamefont {R.}~\bibnamefont {Kueng}}, \bibinfo
  {author} {\bibfnamefont {A.}~\bibnamefont {Neven}}, \bibinfo {author}
  {\bibfnamefont {J.}~\bibnamefont {Carrasco}}, \bibinfo {author}
  {\bibfnamefont {B.}~\bibnamefont {Kraus}}, \bibinfo {author} {\bibfnamefont
  {P.}~\bibnamefont {Zoller}}, \bibinfo {author} {\bibfnamefont
  {P.}~\bibnamefont {Calabrese}}, \bibinfo {author} {\bibfnamefont
  {B.}~\bibnamefont {Vermersch}}, \ and\ \bibinfo {author} {\bibfnamefont
  {M.}~\bibnamefont {Dalmonte}},\ }\href {\doibase
  10.21468/SciPostPhys.12.3.106} {\bibfield  {journal} {\bibinfo  {journal}
  {Scipost Physics}\ }\textbf {\bibinfo {volume} {12}},\ \bibinfo {pages} {106}
  (\bibinfo {year} {2022})}\BibitemShut {NoStop}%
\bibitem [{\citenamefont {Rath}\ \emph {et~al.}(2023)\citenamefont {Rath},
  \citenamefont {Vitale}, \citenamefont {Murciano}, \citenamefont {Votto},
  \citenamefont {Dubail}, \citenamefont {Kueng}, \citenamefont {Branciard},
  \citenamefont {Calabrese},\ and\ \citenamefont
  {Vermersch}}]{rath2023entanglement}%
  \BibitemOpen
  \bibfield  {author} {\bibinfo {author} {\bibfnamefont {A.}~\bibnamefont
  {Rath}}, \bibinfo {author} {\bibfnamefont {V.}~\bibnamefont {Vitale}},
  \bibinfo {author} {\bibfnamefont {S.}~\bibnamefont {Murciano}}, \bibinfo
  {author} {\bibfnamefont {M.}~\bibnamefont {Votto}}, \bibinfo {author}
  {\bibfnamefont {J.}~\bibnamefont {Dubail}}, \bibinfo {author} {\bibfnamefont
  {R.}~\bibnamefont {Kueng}}, \bibinfo {author} {\bibfnamefont
  {C.}~\bibnamefont {Branciard}}, \bibinfo {author} {\bibfnamefont
  {P.}~\bibnamefont {Calabrese}}, \ and\ \bibinfo {author} {\bibfnamefont
  {B.}~\bibnamefont {Vermersch}},\ }\href
  {https://journals.aps.org/prxquantum/abstract/10.1103/PRXQuantum.4.010318}
  {\bibfield  {journal} {\bibinfo  {journal} {PRX Quantum}\ }\textbf {\bibinfo
  {volume} {4}},\ \bibinfo {pages} {010318} (\bibinfo {year}
  {2023})}\BibitemShut {NoStop}%
\bibitem [{\citenamefont {Neven}\ \emph {et~al.}(2021)\citenamefont {Neven},
  \citenamefont {Carrasco}, \citenamefont {Vitale}, \citenamefont {Kokail},
  \citenamefont {Elben}, \citenamefont {Dalmonte}, \citenamefont {Calabrese},
  \citenamefont {Zoller}, \citenamefont {Vermersch}, \citenamefont {Kueng},\
  and\ \citenamefont {Kraus}}]{neven2021symmetryresolved}%
  \BibitemOpen
  \bibfield  {author} {\bibinfo {author} {\bibfnamefont {A.}~\bibnamefont
  {Neven}}, \bibinfo {author} {\bibfnamefont {J.}~\bibnamefont {Carrasco}},
  \bibinfo {author} {\bibfnamefont {V.}~\bibnamefont {Vitale}}, \bibinfo
  {author} {\bibfnamefont {C.}~\bibnamefont {Kokail}}, \bibinfo {author}
  {\bibfnamefont {A.}~\bibnamefont {Elben}}, \bibinfo {author} {\bibfnamefont
  {M.}~\bibnamefont {Dalmonte}}, \bibinfo {author} {\bibfnamefont
  {P.}~\bibnamefont {Calabrese}}, \bibinfo {author} {\bibfnamefont
  {P.}~\bibnamefont {Zoller}}, \bibinfo {author} {\bibfnamefont
  {B.}~\bibnamefont {Vermersch}}, \bibinfo {author} {\bibfnamefont
  {R.}~\bibnamefont {Kueng}}, \ and\ \bibinfo {author} {\bibfnamefont
  {B.}~\bibnamefont {Kraus}},\ }\href {\doibase 10.1038/s41534-021-00487-y}
  {\bibfield  {journal} {\bibinfo  {journal} {npj Quantum Information}\
  }\textbf {\bibinfo {volume} {7}},\ \bibinfo {pages} {152} (\bibinfo {year}
  {2021})}\BibitemShut {NoStop}%
\bibitem [{\citenamefont {Ares}\ \emph
  {et~al.}(2023{\natexlab{a}})\citenamefont {Ares}, \citenamefont {Murciano},\
  and\ \citenamefont {Calabrese}}]{amc-23}%
  \BibitemOpen
  \bibfield  {author} {\bibinfo {author} {\bibfnamefont {F.}~\bibnamefont
  {Ares}}, \bibinfo {author} {\bibfnamefont {S.}~\bibnamefont {Murciano}}, \
  and\ \bibinfo {author} {\bibfnamefont {P.}~\bibnamefont {Calabrese}},\ }\href
  {https://doi.org/10.1038/s41467-023-37747-8} {\bibfield  {journal} {\bibinfo
  {journal} {Nature Communications}\ }\textbf {\bibinfo {volume} {14}},\
  \bibinfo {pages} {2036} (\bibinfo {year} {2023}{\natexlab{a}})}\BibitemShut
  {NoStop}%
\bibitem [{\citenamefont {Ares}\ \emph
  {et~al.}(2023{\natexlab{b}})\citenamefont {Ares}, \citenamefont {Murciano},
  \citenamefont {Vernier},\ and\ \citenamefont {Calabrese}}]{amvc-23}%
  \BibitemOpen
  \bibfield  {author} {\bibinfo {author} {\bibfnamefont {F.}~\bibnamefont
  {Ares}}, \bibinfo {author} {\bibfnamefont {S.}~\bibnamefont {Murciano}},
  \bibinfo {author} {\bibfnamefont {E.}~\bibnamefont {Vernier}}, \ and\
  \bibinfo {author} {\bibfnamefont {P.}~\bibnamefont {Calabrese}},\ }\href@noop
  {} {\  (\bibinfo {year} {2023}{\natexlab{b}})},\ \Eprint
  {http://arxiv.org/abs/2302.03330} {arXiv:2302.03330} \BibitemShut {NoStop}%
\bibitem [{\citenamefont {Bertini}\ \emph {et~al.}(2023)\citenamefont
  {Bertini}, \citenamefont {Klobas}, \citenamefont {Collura}, \citenamefont
  {Calabrese},\ and\ \citenamefont {Rylands}}]{bkccr-23}%
  \BibitemOpen
  \bibfield  {author} {\bibinfo {author} {\bibfnamefont {B.}~\bibnamefont
  {Bertini}}, \bibinfo {author} {\bibfnamefont {K.}~\bibnamefont {Klobas}},
  \bibinfo {author} {\bibfnamefont {M.}~\bibnamefont {Collura}}, \bibinfo
  {author} {\bibfnamefont {P.}~\bibnamefont {Calabrese}}, \ and\ \bibinfo
  {author} {\bibfnamefont {C.}~\bibnamefont {Rylands}},\ }\href@noop {} {\
  (\bibinfo {year} {2023})},\ \Eprint {http://arxiv.org/abs/2306.12404}
  {arXiv:2306.12404} \BibitemShut {NoStop}%
\bibitem [{\citenamefont {Capizzi}\ and\ \citenamefont
  {Mazzoni}(2023)}]{CapizziMazzoni}%
  \BibitemOpen
  \bibfield  {author} {\bibinfo {author} {\bibfnamefont {L.}~\bibnamefont
  {Capizzi}}\ and\ \bibinfo {author} {\bibfnamefont {M.}~\bibnamefont
  {Mazzoni}},\ }\href@noop {} {\enquote {\bibinfo {title} {Entanglement
  asymmetry in the ordered phase of many-body systems: the ising field
  theory},}\ } (\bibinfo {year} {2023}),\ \Eprint
  {http://arxiv.org/abs/2307.12127} {arXiv:2307.12127} \BibitemShut {NoStop}%
\bibitem [{\citenamefont {Fannes}\ \emph {et~al.}(1992)\citenamefont {Fannes},
  \citenamefont {Nachtergaele},\ and\ \citenamefont {Werner}}]{fnw-92}%
  \BibitemOpen
  \bibfield  {author} {\bibinfo {author} {\bibfnamefont {M.}~\bibnamefont
  {Fannes}}, \bibinfo {author} {\bibfnamefont {B.}~\bibnamefont
  {Nachtergaele}}, \ and\ \bibinfo {author} {\bibfnamefont {R.~F.}\
  \bibnamefont {Werner}},\ }\href
  {https://link.springer.com/article/10.1007/BF02099178} {\bibfield  {journal}
  {\bibinfo  {journal} {Communications in Mathematical Physics}\ }\textbf
  {\bibinfo {volume} {144}},\ \bibinfo {pages} {443} (\bibinfo {year}
  {1992})}\BibitemShut {NoStop}%
\bibitem [{\citenamefont {Vidal}(2003)}]{Vidal-03}%
  \BibitemOpen
  \bibfield  {author} {\bibinfo {author} {\bibfnamefont {G.}~\bibnamefont
  {Vidal}},\ }\href
  {https://journals.aps.org/prl/abstract/10.1103/PhysRevLetters91.147902}
  {\bibfield  {journal} {\bibinfo  {journal} {Physical Review Letters}\
  }\textbf {\bibinfo {volume} {91}},\ \bibinfo {pages} {147902} (\bibinfo
  {year} {2003})}\BibitemShut {NoStop}%
\bibitem [{\citenamefont {McCulloch}(2008)}]{mcculloch2008infinite}%
  \BibitemOpen
  \bibfield  {author} {\bibinfo {author} {\bibfnamefont {I.~P.}\ \bibnamefont
  {McCulloch}},\ }\href@noop {} {\enquote {\bibinfo {title} {Infinite size
  density matrix renormalization group, revisited},}\ } (\bibinfo {year}
  {2008}),\ \Eprint {http://arxiv.org/abs/0804.2509} {arXiv:0804.2509}
  \BibitemShut {NoStop}%
\bibitem [{\citenamefont {Vinberg}(1989)}]{Vinberg-89}%
  \BibitemOpen
  \bibfield  {author} {\bibinfo {author} {\bibfnamefont {E.~B.}\ \bibnamefont
  {Vinberg}},\ }\href
  {https://link.springer.com/book/10.1007/978-3-0348-9274-2} {\emph {\bibinfo
  {title} {Linear representations of groups}}}\ (\bibinfo  {publisher}
  {Springer Science \& Business Media},\ \bibinfo {year} {1989})\BibitemShut
  {NoStop}%
\bibitem [{\citenamefont {Ferro}\ \emph {et~al.}(2023)\citenamefont {Ferro},
  \citenamefont {Ares},\ and\ \citenamefont {Calabrese}}]{fac-23}%
  \BibitemOpen
  \bibfield  {author} {\bibinfo {author} {\bibfnamefont {F.}~\bibnamefont
  {Ferro}}, \bibinfo {author} {\bibfnamefont {F.}~\bibnamefont {Ares}}, \ and\
  \bibinfo {author} {\bibfnamefont {P.}~\bibnamefont {Calabrese}},\ }\href@noop
  {} {\  (\bibinfo {year} {2023})},\ \Eprint {http://arxiv.org/abs/2307.06902}
  {arXiv:2307.06902} \BibitemShut {NoStop}%
\bibitem [{\citenamefont {Castro-Alvaredo}\ and\ \citenamefont
  {Doyon}(2011)}]{cd-11}%
  \BibitemOpen
  \bibfield  {author} {\bibinfo {author} {\bibfnamefont {O.~A.}\ \bibnamefont
  {Castro-Alvaredo}}\ and\ \bibinfo {author} {\bibfnamefont {B.}~\bibnamefont
  {Doyon}},\ }\href
  {https://iopscience.iop.org/article/10.1088/1742-5468/2011/02/P02001}
  {\bibfield  {journal} {\bibinfo  {journal} {Journal of Statistical Mechanics:
  Theory and Experiment}\ }\textbf {\bibinfo {volume} {2011}},\ \bibinfo
  {pages} {P02001} (\bibinfo {year} {2011})}\BibitemShut {NoStop}%
\bibitem [{\citenamefont {Castro-Alvaredo}\ and\ \citenamefont
  {Doyon}(2012)}]{cd-12}%
  \BibitemOpen
  \bibfield  {author} {\bibinfo {author} {\bibfnamefont {O.~A.}\ \bibnamefont
  {Castro-Alvaredo}}\ and\ \bibinfo {author} {\bibfnamefont {B.}~\bibnamefont
  {Doyon}},\ }\href {\doibase 10.1103/PhysRevLetters108.120401} {\bibfield
  {journal} {\bibinfo  {journal} {Physical Review Letters}\ }\textbf {\bibinfo
  {volume} {108}},\ \bibinfo {pages} {120401} (\bibinfo {year}
  {2012})}\BibitemShut {NoStop}%
\bibitem [{\citenamefont {Castro-Alvaredo}\ and\ \citenamefont
  {Doyon}(2013)}]{cd-13}%
  \BibitemOpen
  \bibfield  {author} {\bibinfo {author} {\bibfnamefont {O.~A.}\ \bibnamefont
  {Castro-Alvaredo}}\ and\ \bibinfo {author} {\bibfnamefont {B.}~\bibnamefont
  {Doyon}},\ }\href
  {https://iopscience.iop.org/article/10.1088/1742-5468/2013/02/P02016/meta}
  {\bibfield  {journal} {\bibinfo  {journal} {Journal of Statistical Mechanics:
  Theory and Experiment}\ }\textbf {\bibinfo {volume} {2013}},\ \bibinfo
  {pages} {P02016} (\bibinfo {year} {2013})}\BibitemShut {NoStop}%
\bibitem [{\citenamefont {Verstraete}\ and\ \citenamefont
  {Cirac}(2006)}]{vc-06}%
  \BibitemOpen
  \bibfield  {author} {\bibinfo {author} {\bibfnamefont {F.}~\bibnamefont
  {Verstraete}}\ and\ \bibinfo {author} {\bibfnamefont {J.~I.}\ \bibnamefont
  {Cirac}},\ }\href {\doibase 10.1103/PhysRevB.73.094423} {\bibfield  {journal}
  {\bibinfo  {journal} {Physical Review B}\ }\textbf {\bibinfo {volume} {73}},\
  \bibinfo {pages} {094423} (\bibinfo {year} {2006})}\BibitemShut {NoStop}%
\bibitem [{\citenamefont {Hastings}(2007)}]{Hastings-97}%
  \BibitemOpen
  \bibfield  {author} {\bibinfo {author} {\bibfnamefont {M.~B.}\ \bibnamefont
  {Hastings}},\ }\href
  {https://iopscience.iop.org/article/10.1088/1742-5468/2007/08/P08024}
  {\bibfield  {journal} {\bibinfo  {journal} {Journal of statistical mechanics:
  theory and experiment}\ }\textbf {\bibinfo {volume} {2007}},\ \bibinfo
  {pages} {P08024} (\bibinfo {year} {2007})}\BibitemShut {NoStop}%
\bibitem [{\citenamefont {P{\'e}rez-Garc{\'\i}a}\ \emph
  {et~al.}(2008)\citenamefont {P{\'e}rez-Garc{\'\i}a}, \citenamefont {Wolf},
  \citenamefont {Sanz}, \citenamefont {Verstraete},\ and\ \citenamefont
  {Cirac}}]{pwsvc-08}%
  \BibitemOpen
  \bibfield  {author} {\bibinfo {author} {\bibfnamefont {D.}~\bibnamefont
  {P{\'e}rez-Garc{\'\i}a}}, \bibinfo {author} {\bibfnamefont {M.~M.}\
  \bibnamefont {Wolf}}, \bibinfo {author} {\bibfnamefont {M.}~\bibnamefont
  {Sanz}}, \bibinfo {author} {\bibfnamefont {F.}~\bibnamefont {Verstraete}}, \
  and\ \bibinfo {author} {\bibfnamefont {J.~I.}\ \bibnamefont {Cirac}},\ }\href
  {https://journals.aps.org/prl/abstract/10.1103/PhysRevLetters100.167202}
  {\bibfield  {journal} {\bibinfo  {journal} {Physical Review Letters}\
  }\textbf {\bibinfo {volume} {100}},\ \bibinfo {pages} {167202} (\bibinfo
  {year} {2008})}\BibitemShut {NoStop}%
\bibitem [{\citenamefont {Schollwöck}(2011)}]{SCHOLLWOCK201196}%
  \BibitemOpen
  \bibfield  {author} {\bibinfo {author} {\bibfnamefont {U.}~\bibnamefont
  {Schollwöck}},\ }\href {\doibase https://doi.org/10.1016/j.aop.2010.09.012}
  {\bibfield  {journal} {\bibinfo  {journal} {Annals of Physics}\ }\textbf
  {\bibinfo {volume} {326}},\ \bibinfo {pages} {96} (\bibinfo {year}
  {2011})}\BibitemShut {NoStop}%
\bibitem [{\citenamefont {Lieb}\ \emph {et~al.}(1961)\citenamefont {Lieb},
  \citenamefont {Schultz},\ and\ \citenamefont {Mattis}}]{lsm-61}%
  \BibitemOpen
  \bibfield  {author} {\bibinfo {author} {\bibfnamefont {E.}~\bibnamefont
  {Lieb}}, \bibinfo {author} {\bibfnamefont {T.}~\bibnamefont {Schultz}}, \
  and\ \bibinfo {author} {\bibfnamefont {D.}~\bibnamefont {Mattis}},\ }\href
  {\doibase 10.1016/0003-4916(61)90115-4} {\bibfield  {journal} {\bibinfo
  {journal} {Annals of Physics}\ }\textbf {\bibinfo {volume} {16}},\ \bibinfo
  {pages} {407} (\bibinfo {year} {1961})}\BibitemShut {NoStop}%
\bibitem [{\citenamefont {Bonsignori}\ \emph {et~al.}(2023)\citenamefont
  {Bonsignori}, \citenamefont {Capizzi},\ and\ \citenamefont
  {Panopoulos}}]{bcp-23}%
  \BibitemOpen
  \bibfield  {author} {\bibinfo {author} {\bibfnamefont {R.}~\bibnamefont
  {Bonsignori}}, \bibinfo {author} {\bibfnamefont {L.}~\bibnamefont {Capizzi}},
  \ and\ \bibinfo {author} {\bibfnamefont {P.}~\bibnamefont {Panopoulos}},\
  }\href {https://link.springer.com/article/10.1007/JHEP05(2023)027} {\bibfield
   {journal} {\bibinfo  {journal} {Journal of High Energy Physics}\ }\textbf
  {\bibinfo {volume} {2023}},\ \bibinfo {pages} {1} (\bibinfo {year}
  {2023})}\BibitemShut {NoStop}%
\bibitem [{\citenamefont {Molnar}\ \emph {et~al.}(2015)\citenamefont {Molnar},
  \citenamefont {Schuch}, \citenamefont {Verstraete},\ and\ \citenamefont
  {Cirac}}]{msvc-15}%
  \BibitemOpen
  \bibfield  {author} {\bibinfo {author} {\bibfnamefont {A.}~\bibnamefont
  {Molnar}}, \bibinfo {author} {\bibfnamefont {N.}~\bibnamefont {Schuch}},
  \bibinfo {author} {\bibfnamefont {F.}~\bibnamefont {Verstraete}}, \ and\
  \bibinfo {author} {\bibfnamefont {J.~I.}\ \bibnamefont {Cirac}},\ }\href
  {\doibase 10.1103/PhysRevB.91.045138} {\bibfield  {journal} {\bibinfo
  {journal} {Physical Review B}\ }\textbf {\bibinfo {volume} {91}},\ \bibinfo
  {pages} {045138} (\bibinfo {year} {2015})}\BibitemShut {NoStop}%
\bibitem [{\citenamefont {Dubail}(2017)}]{dubail2017entanglement}%
  \BibitemOpen
  \bibfield  {author} {\bibinfo {author} {\bibfnamefont {J.}~\bibnamefont
  {Dubail}},\ }\href
  {https://iopscience.iop.org/article/10.1088/1751-8121/aa6f38/meta} {\bibfield
   {journal} {\bibinfo  {journal} {Journal of Physics A: Mathematical and
  Theoretical}\ }\textbf {\bibinfo {volume} {50}},\ \bibinfo {pages} {234001}
  (\bibinfo {year} {2017})}\BibitemShut {NoStop}%
\bibitem [{\citenamefont {Verstraete}\ and\ \citenamefont
  {Cirac}(2004)}]{vc-04}%
  \BibitemOpen
  \bibfield  {author} {\bibinfo {author} {\bibfnamefont {F.}~\bibnamefont
  {Verstraete}}\ and\ \bibinfo {author} {\bibfnamefont {J.}~\bibnamefont
  {Cirac}},\ }\href@noop {} {\bibfield  {journal} {\bibinfo  {journal} {arXiv
  preprint cond-mat/0407066}\ } (\bibinfo {year} {2004})}\BibitemShut {NoStop}%
\bibitem [{\citenamefont {St{\'e}phan}(2014)}]{Stephan-14}%
  \BibitemOpen
  \bibfield  {author} {\bibinfo {author} {\bibfnamefont {J.-M.}\ \bibnamefont
  {St{\'e}phan}},\ }\href
  {https://iopscience.iop.org/article/10.1088/1742-5468/2014/05/P05010/meta}
  {\bibfield  {journal} {\bibinfo  {journal} {Journal of Statistical Mechanics:
  Theory and Experiment}\ }\textbf {\bibinfo {volume} {2014}},\ \bibinfo
  {pages} {P05010} (\bibinfo {year} {2014})}\BibitemShut {NoStop}%
\bibitem [{\citenamefont {Elben}\ \emph {et~al.}(2023)\citenamefont {Elben},
  \citenamefont {Flammia}, \citenamefont {Huang}, \citenamefont {Kueng},
  \citenamefont {Preskill}, \citenamefont {Vermersch},\ and\ \citenamefont
  {Zoller}}]{elben2023randomized}%
  \BibitemOpen
  \bibfield  {author} {\bibinfo {author} {\bibfnamefont {A.}~\bibnamefont
  {Elben}}, \bibinfo {author} {\bibfnamefont {S.~T.}\ \bibnamefont {Flammia}},
  \bibinfo {author} {\bibfnamefont {H.-Y.}\ \bibnamefont {Huang}}, \bibinfo
  {author} {\bibfnamefont {R.}~\bibnamefont {Kueng}}, \bibinfo {author}
  {\bibfnamefont {J.}~\bibnamefont {Preskill}}, \bibinfo {author}
  {\bibfnamefont {B.}~\bibnamefont {Vermersch}}, \ and\ \bibinfo {author}
  {\bibfnamefont {P.}~\bibnamefont {Zoller}},\ }\href
  {https://www.nature.com/articles/s42254-022-00535-2} {\bibfield  {journal}
  {\bibinfo  {journal} {Nature Reviews Physics}\ }\textbf {\bibinfo {volume}
  {5}},\ \bibinfo {pages} {9} (\bibinfo {year} {2023})}\BibitemShut {NoStop}%
\bibitem [{\citenamefont {Sanz}\ \emph {et~al.}(2009)\citenamefont {Sanz},
  \citenamefont {Wolf}, \citenamefont {P\'erez-Garc\'{\i}a},\ and\
  \citenamefont {Cirac}}]{swpc-09}%
  \BibitemOpen
  \bibfield  {author} {\bibinfo {author} {\bibfnamefont {M.}~\bibnamefont
  {Sanz}}, \bibinfo {author} {\bibfnamefont {M.~M.}\ \bibnamefont {Wolf}},
  \bibinfo {author} {\bibfnamefont {D.}~\bibnamefont {P\'erez-Garc\'{\i}a}}, \
  and\ \bibinfo {author} {\bibfnamefont {J.~I.}\ \bibnamefont {Cirac}},\ }\href
  {\doibase 10.1103/PhysRevA.79.042308} {\bibfield  {journal} {\bibinfo
  {journal} {Physical Review A}\ }\textbf {\bibinfo {volume} {79}},\ \bibinfo
  {pages} {042308} (\bibinfo {year} {2009})}\BibitemShut {NoStop}%
\bibitem [{\citenamefont {Pollmann}\ \emph {et~al.}(2010)\citenamefont
  {Pollmann}, \citenamefont {Turner}, \citenamefont {Berg},\ and\ \citenamefont
  {Oshikawa}}]{ptbo-10}%
  \BibitemOpen
  \bibfield  {author} {\bibinfo {author} {\bibfnamefont {F.}~\bibnamefont
  {Pollmann}}, \bibinfo {author} {\bibfnamefont {A.~M.}\ \bibnamefont
  {Turner}}, \bibinfo {author} {\bibfnamefont {E.}~\bibnamefont {Berg}}, \ and\
  \bibinfo {author} {\bibfnamefont {M.}~\bibnamefont {Oshikawa}},\ }\href
  {\doibase 10.1103/PhysRevB.81.064439} {\bibfield  {journal} {\bibinfo
  {journal} {Physical Review B}\ }\textbf {\bibinfo {volume} {81}},\ \bibinfo
  {pages} {064439} (\bibinfo {year} {2010})}\BibitemShut {NoStop}%
\bibitem [{\citenamefont {Chen}\ \emph {et~al.}(2011)\citenamefont {Chen},
  \citenamefont {Gu},\ and\ \citenamefont {Wen}}]{cgw-11}%
  \BibitemOpen
  \bibfield  {author} {\bibinfo {author} {\bibfnamefont {X.}~\bibnamefont
  {Chen}}, \bibinfo {author} {\bibfnamefont {Z.-C.}\ \bibnamefont {Gu}}, \ and\
  \bibinfo {author} {\bibfnamefont {X.-G.}\ \bibnamefont {Wen}},\ }\href@noop
  {} {\bibfield  {journal} {\bibinfo  {journal} {Phys. Rev. B}\ }\textbf
  {\bibinfo {volume} {84}},\ \bibinfo {pages} {235128} (\bibinfo {year}
  {2011})}\BibitemShut {NoStop}%
\end{thebibliography}%

\appendix

\section{Symmetrization of a density matrix}\label{app:symm}

In this appendix, we explain how to construct explicitly the symmetrized state $\tilde{\rho}_A$ in terms of the block decomposition of $\rho_A$ wrt the symmetry sectors, making contact with the original definition of Ref.~\cite{amc-23} proposed for $U(1)$. We first summarize the main result, and then we discuss its derivation. 

Let us decompose the Hilbert space $\mathcal{H}_A$ in sectors, say irreducible representations of $G$, as follows
\be
\mathcal{H}_A = \oplus_{\sigma} \oplus^{n_\sigma}_{j=1} V_{\sigma,j}.
\ee
Here $\sigma$ labels a generic irreducible representation of $G$, and $V_{\sigma,j}\subset \mathcal{H}_A$ are $n_\sigma$ copies of it. For any operator acting on $\mathcal{H}_A$ we have a block decomposition given by the matrix elements connecting two generic subspace $V_{\sigma,j}$ and $V_{\sigma',j'}$. Here, we state that
\begin{itemize}
\item The blocks of $\tilde{\rho}_A$ connecting two distinct irreducible representations $\sigma$ and $\sigma'$ are vanishing.
\item The block of $\tilde{\rho}_A$ connecting $V_{\sigma,j}$ and $V_{\sigma,j'}$ is proportional to the identity. Moreover, its trace is equal to the corresponding block of $\rho_A$, a property which fixes unambiguously the proportionality constant.
\end{itemize}

\begin{figure}
    \centering
    \includegraphics[width=\linewidth]{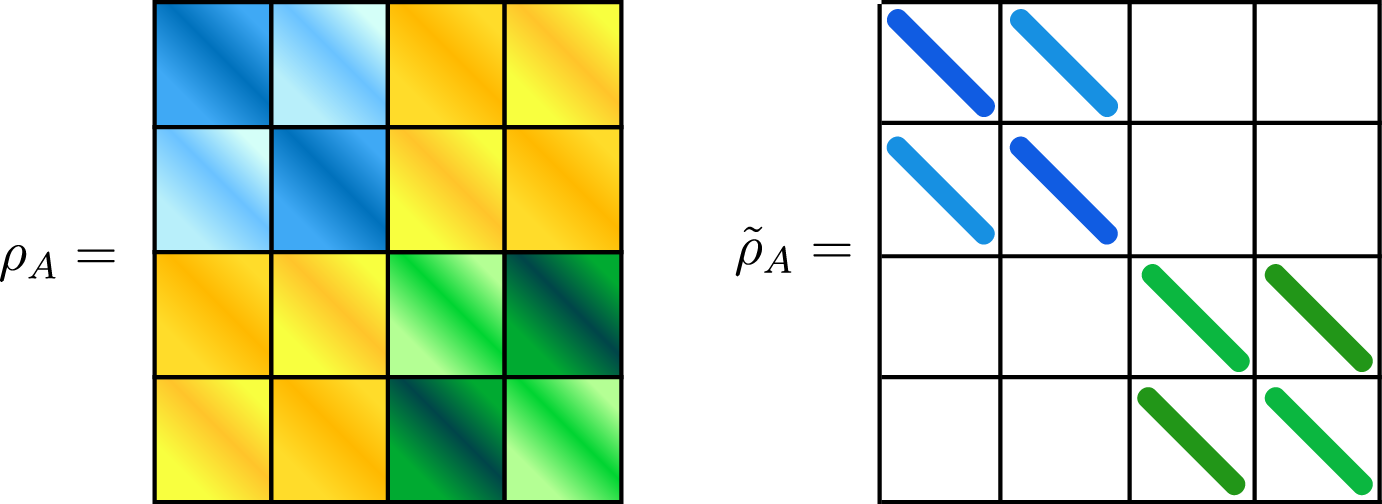}
    \caption{Symmetrization of the reduced density matrix. 
    The blocks that connect equivalent representations, the blue and green ones, become proportional to the identity. The other (yellow) ones are washed out by the symmetrization procedure.}
    \label{fig:sym_RDM}
\end{figure}
This is represented pictorially in Fig. \ref{fig:sym_RDM}. The result above is a direct consequence of the well-known Schur lemma, and we refer the reader to Ref.~\cite{Vinberg-89} for details. For completeness, we discuss first a simple derivation for abelian groups, and then we provide a complementary approach for generic (non-abelian) groups.

\subsection{Abelian groups}

For abelian groups, a straightforward calculation starting from the definition leads directly to the result. 

We first observe that, since every irreducible representation is one-dimensional (see Ref.~\cite{Vinberg-89}), that is $\text{dim}\l V_{\sigma,j}\r=1$, $\hat{g}_A$ is proportional to the identity on the subspace $\oplus_j V_{\sigma,j}$. In particular, given $\chi_\sigma$ the character of $\sigma$ and $\Pi_\sigma$ the orthogonal projector onto $\oplus_j V_{\sigma,j}$, it holds
\be\label{eq:g_char}
\hat{g}_A = \sum_\sigma \chi_\sigma(g)\Pi_\sigma.
\ee
Inserting the result above onto the definition \eqref{eq:rho_tilde} we get
\be\label{eq:rho_tilde_sectors}
\begin{split}
\tilde{\rho}_A = & \frac{1}{|G|}\sum_{g,\sigma,\sigma'} \chi_\sigma(g)\overline{\chi_{\sigma'}(g)} \Pi_\sigma \rho \Pi_{\sigma'} =\\
&\sum_\sigma \Pi_\sigma \rho \Pi_\sigma,
\end{split}
\ee
where the orthogonality of characters \cite{Vinberg-89}
\be
\frac{1}{|G|}\sum_g \chi_\sigma(g)\overline{\chi_{\sigma'}(g)} = \delta_{\sigma \sigma'}
\ee
has been employed. We stress that Eq.~\eqref{eq:rho_tilde_sectors}, proved here for finite abelian groups, holds for abelian compact Lie groups too.

In contrast, Eq.~\eqref{eq:rho_tilde_sectors} does not apply generically to non-abelian groups. The reason is that for those groups a direct relation as Eq.~\eqref{eq:g_char} between the action of the group elements and the characters is absent. Specifically, whenever an irreducible representation of dimension greater than $1$ appears in $\mathcal{H}_A$, then Eq.~\eqref{eq:g_char} does not hold.

It is also worth to explain why 
Eq.~\eqref{eq:rho_tilde_sectors} cannot hold on physical ground for a non-abelian group as $SU(2)$. Imagine, for example, that $\rho_A$ has a block of spin $S>0$ with distinct entries along the diagonal, corresponding to distinct probabilities to observe the values of $S_z$, the magnetization along the $z$ axis. The same property is clearly shared by the matrix $\sum_\sigma \Pi_\sigma \rho_A \Pi_\sigma$, as its entries on the block above are the same as $\rho_A$. However, $\tilde{\rho}_A$ is symmetric under rotation by construction, which implies that the probabilities associated with $S_z$ cannot depend on the explicit value of the magnetization. This observation clearly rules out Eq.~\eqref{eq:rho_tilde_sectors}, and suggests the properties the blocks of $\tilde{\rho}_A$ should have.

\subsection{General case}

To understand what happens in the general case, it is convenient to think $\text{End}\l \mathcal{H}_A\r$ as a Hilbert space with the Hilbert Schmidt product $\la f_1,f_2\ra \equiv \text{Tr}\l {f_1}^\dagger f_2\r$. Within this perspective, $\rho_A \rightarrow \tilde{\rho}_A$ can be seen as an orthogonal projection onto the subspace of symmetric operators
\be
\text{End}_G(\mathcal{H}_A) \equiv \{f \in  \text{End}(\mathcal{H}_A) \ | \ [\hat{g}_A,f] =0  \ \forall g \in G \}.
\ee
In this way, the problem of the symmetrization is traced back to find an orthogonal basis of $\text{End}_G(\mathcal{H}_A)$ and project $\rho_A$ onto it.

First, the Schur lemma ensures that any symmetric operator connecting two distinct irreducible representations has to vanish. Furthermore, it also tells us that $n_\sigma^2$ independent invariant operators acting on the sector of $\sigma$ are present, a relation that we write as
\be
\text{dim}\l  \text{End}_G\l \oplus^{n_\sigma}_{j=1} V_{\sigma,j} \r\r = n_\sigma^2.
\ee
To construct those operators explicitly, we choose an orthonormal basis of $V_{\sigma,j}$ as
\be
\{\ket{\sigma,j,a}\}, \quad a=1,\dots,\text{dim}(\sigma),
\ee
with $j=1,\dots,n_\sigma$. Then, it is easy to show that
\be
I_{\sigma,jj'} \equiv \frac{1}{\sqrt{\text{dim}(\sigma)}}\sum^{\text{dim}(\sigma)}_{a=1} \ket{\sigma,j,a} \bra{\sigma,j',a},
\ee
is a symmetric operator connecting $V_{\sigma,j'}$ with $V_{\sigma,j}$, whose corresponding block is proportional to the identity.

In conclusion, the operators $\{I_{\sigma,jj'}\}_{j,j'=1,\dots,n_\sigma}$, which are independent and orthogonal by construction, constitute an orthonormal basis for $\text{End}_G(\mathcal{H}_A)$, and one eventually expresses
\be
\tilde{\rho}_A =\sum_\sigma \sum^{n_\sigma}_{j,j'=1} \la I_{\sigma,jj'},\rho_A\ra I_{\sigma,jj'}.
\ee

\section{Area law saturation and exponential correction}\label{app:Area_law}

Here, we comment on the exponential corrections to the area-law, that is Eq.~\eqref{eq:area_law},  showing that they have a different origin wrt the exponential decay of the charged moments in Eq.~\eqref{eq:ch_mom_decay}.

To compute those corrections, it is sufficient to expand $R$ in terms of its spectrum as
\be\label{eq:R_expansion}
R \simeq \Pi + \lambda \Pi' +\dots,
\ee
with $|\lambda|<1$ the next-to-leading eigenvalue and $\Pi'$ its corresponding projector (satisfying $\Pi'\Pi=0$ and $(\Pi')^2 = \Pi'$). In the large $|A|$ limit, from Eq.~\eqref{eq:R_expansion} one gets
\be
R^{|A|} \simeq \Pi + \lambda^{|A|} \Pi' + \dots,
\ee
and similarly
\be\label{R_expansion1}
\begin{split}
(R^{\otimes n})^{|A|} \simeq \Pi^{\otimes n} + \lambda^{|A|}( \Pi'\otimes \Pi\dots \otimes \Pi + \\
\Pi\otimes \Pi' \dots \otimes \Pi + \dots ) +\dots .
\end{split}
\ee
Inserting Eq.~\eqref{R_expansion1} in Eq.~\eqref{eq:Renyi_MPS} one finally gets
\be
\begin{split}
\text{Tr}\l \rho_A^n\r \simeq &\text{Tr}\l \Pi^{\otimes n} P \Pi^{\otimes n} P^\dagger \r +\\
&n \lambda^{|A|} \text{Tr}[ (\Pi'\otimes \Pi^{\otimes n}) P \Pi^{\otimes n} P^\dagger ] + \dots,
\end{split}
\ee
where we used that the $n$ terms of Eq.~\eqref{R_expansion1} give the same contribution, due to the symmetric properties of the operator $P$ under replica shift.

Similar considerations hold for the charged moments of symmetric elements, namely belonging to $H$ (see Eq.~\eqref{eq:area_law1}). In particular, we show below that the decay rate of their exponential corrections is the same as the ones for the R\'enyi entropy. We employ the well-known result that the state Eq.~\eqref{eq:PsiL} is invariant under $g$, say $g\in H$, iff the tensor $M$ is invariant under $u_g$ up to a change of basis and a phase, say (see Ref. \cite{swpc-09,ptbo-10})
\be
\sum_{s'} (M_{s'})_{a,b} (u_{g})_{ss'} = e^{i \phi} \sum_{b'a'} U_{b'b}(M_{s})_{ab}(U^{-1})_{aa'} ,
\ee
with $U$ an invertible $D\times D$ matrix and $\phi \in \mathbb{R}$. Therefore, it is evident that whenever $g\in H$, the eigenvalues of $R_g$ are related to the one of $R$ via the phase-shift $e^{i\phi}$. In particular, the next-to-leading eigenvalue of $R_g$ has the same absolute value of $\lambda$.

\section{Clustering hypothesis}\label{app:clust_hyp}

An important hypothesis underlying our derivation is the clustering property of the state under analysis, say
\be
\la \mathcal{O}(\mathbf{x})\mathcal{O}(\mathbf{y})\ra \rightarrow \la \mathcal{O}(\mathbf{x})\ra \la\mathcal{O}(\mathbf{y})\ra, \quad |\mathbf{x}-\mathbf{y}|\rightarrow \infty
\ee
with $\mathcal{O}(\mathbf{x}),\mathcal{O}(\mathbf{y})$ local observables with support localized around the sites $\mathbf{x},\mathbf{y}$ respectively. This is technically equivalent to non-degeneracy of the largest eigenvalue of $R$ in Eq. \eqref{eq:Rdef}, as explained in Ref. \cite{cgw-11}. In this appendix, we show that the clustering hypothesis is crucial for the validity of our main result, and we discuss a simple paradigmatic counterexample.

Let us consider the state of a spin-1/2 chain
\be
\ket{\Psi} = \sqrt{p}\ket{\uparrow \dots \uparrow} + \sqrt{1-p}\ket{\downarrow \dots \downarrow},
\ee
with $0\leq p\leq 1$. Since $\ket{\Psi}$ is a linear combination of product states, it can be realized as an MPS (of dimension $2$). Also $\ket{\Psi}$ is not clustering, as it holds
\be
\la \sigma^z_x \sigma^z_y\ra = 1, \quad x\neq y
\ee
which differs explicitly from
\be
\la \sigma^z_x\ra \la \sigma^z_y\ra = (2p-1)^2
\ee
whenever $p\neq 0,1$.

We compute the reduced density matrix of $\ket{\Psi}$ of a (proper) subsystem as
\be
\rho_A = p\ket{\uparrow \dots \uparrow}\bra{\uparrow \dots \uparrow} +  (1-p)\ket{\downarrow \dots \downarrow}\bra{\downarrow \dots \downarrow},
\ee
so that its symmetrization under $\mathbb{Z}_2$, induced by the spin-flip operator $\sigma^x$, is
\be
\tilde{\rho}_A = \frac{1}{2}\la \ket{\uparrow \dots \uparrow}\bra{\uparrow \dots \uparrow} +\ket{\downarrow \dots \downarrow}\bra{\downarrow \dots \downarrow}\ra.
\ee
Therefore, the R\'enyi entanglement asymmetry of $\ket{\Psi}$ is
\be\label{eq:DS_counterex}
\Delta S_n = \log 2 - \frac{1}{1-n}\log\l p^n+(1-p)^n\r,
\ee
no matter the size of the subsystem $A$. We remark that for $p=0,1$ the state is clustering,it breaks $\mathbb{Z}_2$, and the result $\Delta S_n =\log 2$ is compatible with our universal prediction \eqref{eq:conj}.
Moreover, for $p=1/2$ one has $\Delta S_n=0$, which is still compatible with \eqref{eq:conj}, as the $\mathbb{Z}_2$ symmetry is unbroken, but it does not follow from our derivation since the clustering hypothesis does not hold. In all the other cases, Eq. \eqref{eq:DS_counterex} violates explicitly the prediction \eqref{eq:conj}.

Finally, it is worth noting that this simple example leads to an important remark in the context of spontaneous symmetry breaking. For instance, one can straightforwardly employ our predictions to the ground states with short-range correlations, say the ground states of an Ising with definite magnetization, but violations are expected to appear whenever linear combinations of those are taken. In particular, depending on the boundary conditions of the Hamiltonian, a non-clustering ground state can be selected, as it happens for the Ising model for periodic boundary conditions; consequently, one should be careful in those cases, and tune properly the boundary terms to select correctly a clustering ground state.

\section{Saddle point analysis of the charged moments}\label{app:saddle}

Here, we give some details and comment on some subtleties regarding the saddle point analysis employed in Sec. \ref{sec:Lie}. The crucial quantity in the forthcoming discussion is the function $f(\mathbf{g})$, defined in Eq.~\eqref{eq:f_fun} and its behavior in the large volume limit $|A|\rightarrow \infty$.

First, we observe that, thanks to the analysis of \ref{sec:fin_groups} (see e.g. Eq.~\eqref{eq:ch_mom_decay}), the limit above exists
\be
F(\mathbf{g}) \equiv -\underset{|A|\rightarrow \infty}{\lim}\frac{1}{|A|}\log f(\mathbf{g}),
\ee
and we refer to it as \textit{density of charged free energy}. Also, $F(\mathbf{g})=0$ iff $\mathbf{g} \in H^{n-1}$ and it is a continuous function of its entry. In general, $F(\mathbf{g})$ is not guaranteed to be smooth: indeed, as its value is related to the largest eigenvalue of the matrix $R_g$, singularities might appear $R_{g}$ displays level crossing. However, we are only interested in the leading asymptotic of the integral Eq.~\eqref{eq:integ_f}, and an analysis of $F(\mathbf{g})$ in a neighborhood of $H^{n-1}$ is sufficient for our purpose. For instance, we aim to prove that
\begin{itemize}
\item $F(\mathbf{g})$ is analytic in a neighborhood of $H^{n-1}$.
\item $H^{n-1}$ is a manifold of saddle points for $F(\mathbf{g})$ and Eq.~\eqref{eq:f_2ord} holds.
\end{itemize}

The first property is a consequence of the assumption that $\ket{\Psi}$ satisfies clustering. Indeed, for $g=1$ the largest eigenvalue of $R_g = R$ is gapped: this implies that the largest eigenvalue of $R_g$ is a smooth function in a neighborhood of $g=1$, as it does not cross other eigenvalues. The same considerations hold for $g$ belonging to a neighborhood of $H$, as one can show that $R_g$ has the same spectrum of $R$ up to an overall phase whenever $g \in H$ and it is a symmetry of the state (see \cite{swpc-09}).

To prove the second property, we have to show that no linear terms appear whenever $F(\mathbf{g})$ is Taylor expanded around $\mathbf{g} \in H^{n-1}$ and that the quadratic form in Eq.~\eqref{eq:f_2ord} is non-degenerate. Expanding $f(\mathbf{g})$ in Eq.~\eqref{eq:f_fun} at first order near $\mathbf{g} \in H^{n-1}$ we get
\be
\begin{split}
&f(\mathbf{g}) \simeq \frac{\text{Tr}(\rho_A h_1(1+X_1)\rho_A h_2 \dots)}{\text{Tr}(\rho^n_A)}+\dots+\\
&\frac{\text{Tr}(\rho_A h_1\rho_A h_2 \dots \rho_A (1-X_{n-1})h^{-1}_{n-1}\dots (1-X_{1})h^{-1}_{1}) }{\text{Tr}(\rho^n_A)}\simeq 1,
\end{split}
\ee
where the cyclicity of the trace, together with $[h_j,\rho_A] =0$ has been used. Therefore, as no linear terms appear in $f(\mathbf{g})$, the same holds for $F(\mathbf{g})$.

While we did not find rigorous proof of the non-degeneracy of the quadratic form in Eq.~\eqref{eq:f_2ord}, we expect that a hypothetical violation can only be a fine-tuning of the model. Indeed, even if in principle it might be possible that $F(\mathbf{g})$ vanishes faster than $O(X^2)$ along an infinitesimal curve that is not tangent to $H^{n-1}$, and, e.g. the leading term is at order $O(X^4)$, we were not able to find an explicit example where this scenario occurs.

\end{document}